# Indistinguishable binomial decision tree of 3-SAT: Proof of P $\subsetneq$ NP

Keum-Bae Cho

**Affiliations:**

B.S. in Electrical Engineering at KAIST, South Korea; M.S. and Ph.D. in Electrical Engineering and Computer Science at Seoul National University, South Korea

*Corresponding author, E-mail (kbcho98@snu.ac.kr)

**Abstract**: This paper solves a long standing open problem of whether NP-complete problems could be solved in polynomial time on a deterministic Turing machine by showing that the indistinguishable binomial decision tree can be formed in a 3-SAT instance. This paper describes how to construct the decision tree and explains why 3-SAT has no polynomial-time algorithm when the decision tree is formed in the 3-SAT instance. The indistinguishable binomial decision tree consists of polynomial numbers of nodes containing an indistinguishable variable pair but generates exponentially many paths connecting the clauses to be used for sequences of resolution steps. The number of paths starting from the root node and arriving at a child node forms a binomial coefficient. In addition, each path has an indistinguishable property from one another. Due to the exponential number of paths and their indistinguishability, if an indistinguishable binomial decision tree is constructed in which there exist one or more paths generating an empty clause, the number of calculation steps needed to extract the empty clause is not polynomially bounded. This result leads to the conclusion that class P is a proper subset of class NP.

**One Sentence Summary:** This paper solves the P versus NP problem by showing that the indistinguishable binomial decision tree can be formed in a 3-SAT instance.

## INTRODUCTION

The Boolean satisfiability (SAT) problem is to determine whether there exists a feasible set to satisfy a given Boolean formula. SAT is the first known example of a NP-complete problem[2] (Cook–Levin theorem) and thousands of NP-compete problems have been identified by reducing the SAT to the problems. The SAT problem is divided by tractable SAT such as 2-SAT and Horn-SAT and intractable SAT such as 3-SAT[3]. In the above tractable instances, 2-SAT is NL-complete[4] and Horn-SAT is P-complete[5]. In addition, 3-SAT is NP-complete[2]. Hence, SAT is a good research object to search for the relationship of the classes, NL, P, and NP. It is known that NL $\subseteq$ P $\subseteq$ NP, but unknown whether NL=P and whether P=NP. These two questions have been open problems for several decades. Most complexity theorists expect that NL $\subsetneq$ P $\subsetneq$ NP. Especially, the P versus NP problem[6], as one of the famous unsolved problems in mathematics and computer science, is to clarify the relationship for the inclusion of the classes P and NP[7]. The obvious way to prove P = NP is to show that one or more NP-complete problems have a polynomial-time algorithm. Researchers have found thousands of NP-complete problems since Karp's research[8,9]. However, although there are so many NP-complete problems, researchers have failed to find a polynomial-time algorithm for any of the problems. Hence, with the belief that $P \neq NP$, various proof techniques have been studied to distinguish between P and NP. However, all known proof techniques such as relativizing[10], natural[11~16], and algebrizing[17~20] proofs were insufficient to prove that $P \neq NP$. Resolving the question whether 3-SAT has a polynomial-time algorithm is equivalent to the P versus NP problem because 3-SAT is NP-complete. This paper solves the P versus NP problem by showing that 3-SAT has no polynomial-time algorithm. The SAT problem is organized by clauses. The simplest clause is the unit clause containing only one variable. This paper shows that the simplest clause can be converted to a logically equivalent, highly complicated clause group, which is termed as a binomial decision tree. The decision tree has polynomial number of nodes but generates exponentially many paths arriving at the polynomial number of nodes. One of the distinctive features of the indistinguishable binomial decision tree is the indistinguishability of paths in that the paths cannot be divided into groups according to the arrival node. As the paths cannot be grouped, every algorithm must search for all exponential number of paths in order to verify whether there exists a node containing a specific variable.

The P versus NP problem is explained as to whether every problem whose solution can be quickly (in polynomial time) verified, can also be solved quickly. We will show that every algorithm must investigate all nodes to decide the satisfiability of an instance. An instance containing the binomial decision tree is quickly verifiable because the number of nodes is polynomially bounded. However, to solve the problem, that is, to investigate all nodes cannot be executed quickly because exponentially many paths must be investigated to decide the satisfiability. Therefore, the indistinguishable binomial decision tree clearly explains the relationship of verifiability and solvability of NP-complete problems. This paper describes how to construct the indistinguishable binomial decision tree.

# RESULTS

### Decision chain and generalized unit clause

In order to verify the satisfiability of an instance represented by conjunctive normal form (CNF), we investigate whether we can generate an empty clause from some set of clauses. In a logical approach, the only way to generate a logically equivalent new clause is to apply the resolution rule. For instance, there is a resolution step to apply the resolution rule:

$$\frac{a \vee c, \quad b \vee \neg c}{a \vee b} \quad \cdots(1)$$

The dividing line stands for entails, which means that $(a \vee c) \wedge (b \vee \neg c)$ is logically equivalent to $(a \vee c) \wedge (b \vee \neg c) \wedge (a \vee b)$. The input clauses must have a variable and its complement, which is called as resolved variable. We use the concept of the **decision chain**[1] to represent a sequence of resolution steps. The decision chain is a linked list of the resolved variables while sequentially applying the resolution rule. Figure 1 shows an example of a sequence of resolution steps and its representation by a decision chain.

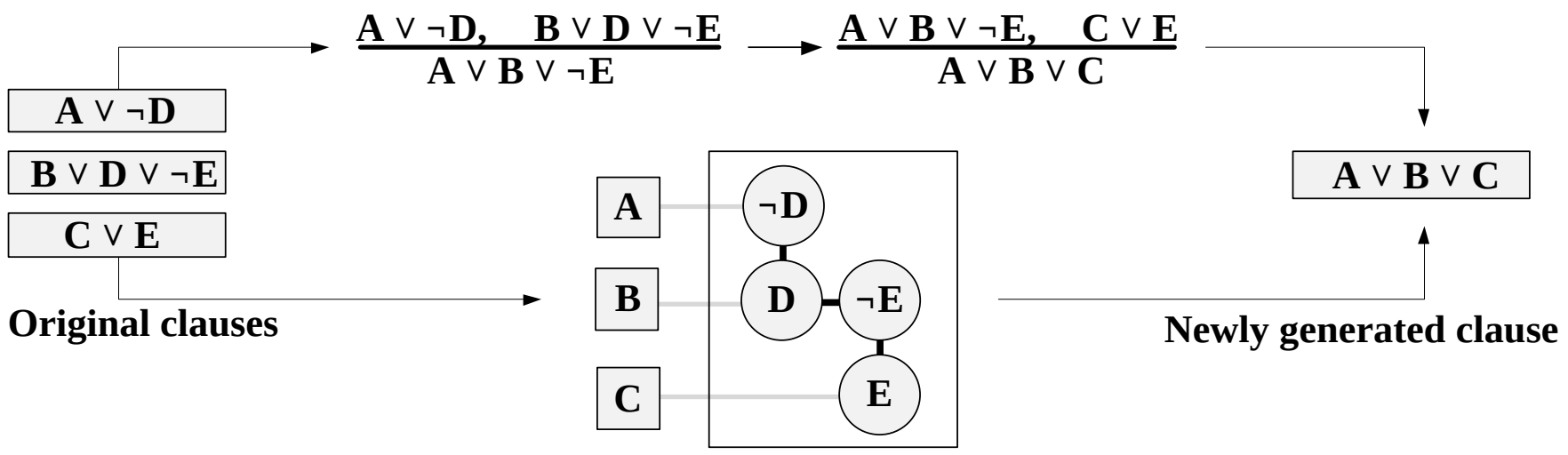

**Fig. 1.** A sequence of resolution steps and its representation by a decision chain

The decision chain shows the whole process of resolution steps with a linked list using the graphical representation. The decision chain connected with $n$ ($n \geq 0$) number of variables is logically equivalent to a clause containing $n$ variables because the decision chain is a linked list of resolved variables, which correspond to removed variables in newly generated clauses. Therefore, a decision chain that is not connected with any variable is logically equivalent to an empty clause, which is termed as a **generalized empty clause.** In addition, a decision chain connected with only one variable is logically equivalent to a unit clause, which is termed as a **generalized unit clause**.

If a sequence of resolution steps generates an empty clause from some set of clauses in an instance, the instance is unsatisfiable. Likewise, if one or more generalized empty clauses are constructed with some set of clauses in an instance, the instance is unsatisfiable. The process building a decision chain to construct a generalized unit clause or generalized empty clause corresponds to the process searching of clauses for a sequence of resolution steps to generate a unit clause or empty clause.

A variable contained in a unit clause must be assigned with only one value of ‘0’ or ‘1’ to satisfy the clause, which is termed as a **dominant variable**. Likewise, a variable connected alone to a decision chain becomes a dominant variable, which must be assigned with only one value of ‘0’ or ‘1’ to satisfy all clauses contained in the decision chain. We say that every dominant variable has **dominance property**. For example, variable $x$ in a unit clause is a dominant variable and the variable has dominance property. If ‘$x$’ has dominance property and ‘$\neg x$’ also has dominance property, which means that ‘$x$’ is connected alone to a decision chain and ‘$\neg x$’ is also connected alone to another decision chain, then the instance is unsatisfiable. If two or more variables are connected to a decision chain, the variables are termed as dominant variable candidates.

**The necessity of searching for the unit clause**

Algorithms for tractable SAT such as 2-SAT and Horn-SAT are based on the unit propagation technique, which starts in the search for unit clauses. If some algorithm cannot find the unit clause in a CNF instance, is it possible to solve a SAT problem with the algorithm?

A variable in a clause must be a dominant variable only when all the other variables contained in the clause are dominant variables and their feasible values do not satisfy the clause. Otherwise, the variable can be assigned with any value of ‘0’ or ‘1’. Hence, the feasible value of a variable cannot be decided without investigating the other variable’s dominance property. Suppose that we decide the feasible value of a variable at first. We cannot decide the feasible value because we do not know the other variable’s dominance property. The feasible value of a variable can be decided only when the variable has no relationship with the other variables in a clause. The variable contained in the unit clause satisfies this condition. Therefore, we must search for a unit clause to decide the satisfiability of an instance. However, we can construct an instance that contains no unit clauses but contains generalized unit clauses.

A generalized unit clause is constructed by transforming a unit clause into a decision chain. This transformation makes it hard to find a sequence of resolution steps to generate the unit clause when the decision chain is extended to a tree named as a binomial decision tree. For this transformation, we newly introduce the concept of the indistinguishable variable pair and decision tree generating indistinguishable paths. We show that if we construct a binomial decision tree in which every clause contains an indistinguishable variable pair, then the number of paths to reach all clauses of the tree is not polynomially bounded. In addition, due to the indistinguishability of paths, we must search for all exponential number of paths to investigate all polynomial number of clauses, which requires exponentially many calculation steps. First, we introduce how to transform a unit clause into a decision chain.

**Unit clause transformation by indistinguishable variable pairs**

In order to generate a hard SAT instance, we transform a unit clause containing variable $x_1$ into a decision chain with $k$ number of clauses consisting of two variables. The unit clause containing $x_1$ is represented with a logically equivalent CNF consisting of $k$ numbers of clauses such that:

$$x_1 \equiv (x_1 \vee x_2) \wedge (\neg x_2 \vee x_3) \wedge (\neg x_3 \vee x_4) \wedge \ldots \wedge (\neg x_{k-1} \vee x_k) \wedge (\neg x_k \vee x_1) \qquad \cdots(2)$$

If we sequentially execute resolution steps $k$-1 times with the clauses in eq. (2), a new clause ($x_1 \vee x_1$) is generated, with which we can verify the dominance property of $x_1$.

In order to increase the hardness of the instance, we divide variable $x_i$ ($2 \le i \le k$) with $x_{i.1}$ and $x_{i.2}$, and we replace $\neg x_i$ with $\neg x_{i.1}$. In addition, we add two clauses ($\neg x_{i-1.1} \vee x_{i.1} \vee \neg x_{i.2}$) and ($\neg x_{i-1.1} \vee \neg x_{i.1} \vee x_{i.2}$) in the clause set of an instance, with which two clauses ($\neg x_{i-1.1} \vee x_{i.1}$) or ($\neg x_{i-1.1} \vee x_{i.2}$) can be generated by one resolution step. Through the above conversion, a resolution step to generate clause ($x_1 \vee x_3$) such that

$$\frac{x_1 \vee x_2,\ \neg x_2 \vee x_3}{x_1 \vee x_3} \quad \cdots(3)$$

is converted to a sequence of resolution steps to generate clause ($x_{1.1} \vee x_{3.1}$) such that

$$\frac{x_{1.1} \vee x_{2.1} \vee x_{2.2},\ x_{1.1} \vee x_{2.1} \vee \neg x_{2.2}}{x_{1.1} \vee x_{2.1}},\ \frac{\neg x_{2.1} \vee x_{3.1} \vee x_{3.2},\ \neg x_{2.1} \vee x_{3.1} \vee \neg x_{3.2}}{\neg x_{2.1} \vee x_{3.1}},\ \frac{x_{1.1} \vee x_{2.1},\ \neg x_{2.1} \vee x_{3.1}}{x_{1.1} \vee x_{3.1}} \quad \cdots(4)$$

In this case, we need to select one variable between the two twice. We cannot distinguish two variables because two variables are divided from one variable. Hence, we must arbitrarily select one variable. Two divided variables are termed as an **indistinguishable variable pair**. Figure 2 shows the process of transforming a unit clause by indistinguishable variable pairs

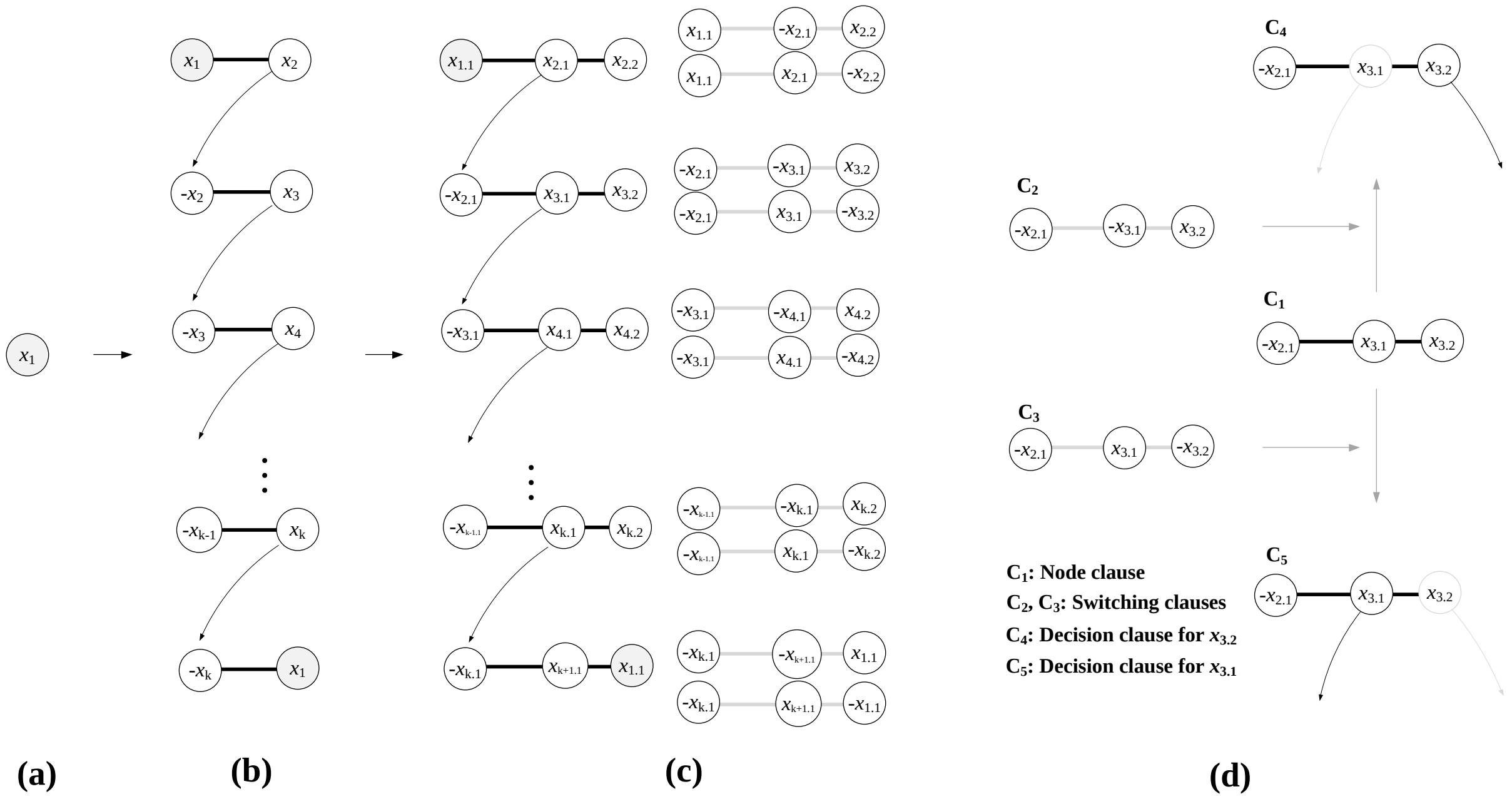

**Fig. 2.** Unit clause transformation into a decision chain consisting of indistinguishable variable pairs

We used notation **'-*x*'** instead of **'¬*x*'** in Figure 2. Figure 2(a) and Figure 2(b) represent a unit clause and a decision chain transformed from the unit clause, respectively. The CNF corresponding to the decision chain is represented by eq. (2). Figure 2(c) represents a decision chain consisting of indistinguishable variable pairs. The CNF corresponding to the decision chain of Figure 2(c) is represented as

$$(x_{1.1} \vee x_{2.1} \vee x_{2.2}) \wedge (x_{1.1} \vee x_{2.1} \vee -x_{2.2}) \wedge (-x_{2.1} \vee x_{3.1} \vee x_{3.2}) \wedge (-x_{2.1} \vee x_{3.1} \vee -x_{3.2}) \wedge \ldots \wedge (-x_{k.1} \vee x_{k+1.1} \vee x_{1.1}) \wedge (-x_{k.1} \vee -x_{k+1.1} \vee x_{1.1}) \quad \cdots(5)$$

We can easily verify that eq. (5) is logically equivalent to $x_{1.1}$. We replaced a clause consisting of two variables with two clauses consisting of three variables such that one is a clause containing two indistinguishable variables, which is termed as a **node clause**, and another is an added clause to remove one variable, which is termed as a **switching clause**. It is not necessary to restrict the switching clause as the clause containing three variables such as ($\neg x_{2.1} \vee \neg x_{3.1} \vee x_{3.2}$) and ($\neg x_{2.1} \vee$ $x_{3.1} \vee \neg x_{3.2}$). The switching clause can be ($\neg x_{3.1} \vee x_{3.2}$) and ($x_{3.1} \vee \neg x_{3.2}$), and any polynomial number of clauses generating ($\neg x_{3.1} \vee x_{3.2}$) and ($x_{3.1} \vee \neg x_{3.2}$) by a sequence of resolution steps. The decision chain consists of clauses containing two variables. If a node clause can be replaced with a clause consisting of two variables by removing one indistinguishable variable, the node clause is termed as a **decision clause** for the remaining indistinguishable variable. The path generated by sequentially connected decision clauses is termed as a **decision path**. We can directly insert the switching clauses in the clause set of an instance. In this case, the decision clause is termed as an **explicit decision clause**. In addition, we can construct decision paths to generate switching clauses instead of the direct insertion. In this case, the decision clause is termed as an **implicit decision clause**. Figure 2(d) shows an example that a node clause plays a role of a decision clause by two switching clauses. If a node clause cannot become a decision clause for an indistinguishable variable, all resolvent clauses generated by a sequence of resolution steps using the node clause and the other clauses sequentially connected to the variable, contain at least two variables. Hence, if we are searching for a decision path via a node clause to investigate whether a unit clause can be generated by a sequence of resolution steps, we must verify whether the node clause becomes a decision clause in advance. For this verification, we must search for switching clauses. Hereafter, we omit switching clauses in drawing the chain and tree.

**Binary decision tree of 3-SAT**

Suppose that one variable is selected from the indistinguishable variable pair in a node. If an algorithm terminates after executing several resolution steps without encountering any branching, we can select another variable after returning to the previous node. For example, suppose that clauses ($\neg x_{i.2} \vee \neg x_{1.1}$) ($2 \le i \le k$) are added to the clause set of an instance. In this case, if we choose $x_{i.2}$, one resolution step generates a clause ($x_{1.1} \vee \neg x_{1.1}$). There is no variable to continue the resolution steps any more. If we go back and select another variable, only several steps are added in the number of calculation steps, which does not affect the calculation complexity of the algorithm. Hence, In order to construct harder SAT instance, we construct an instance in which two nodes connected with two indistinguishable variables have the same depth. In order to satisfy this condition in all nodes, the decision chain must be extended to a tree as shown in Figure 3, which is termed as a **decision tree**. Note that the decision tree consists of decision clauses. That is, all node clauses are an explicit decision clause or an implicit decision clause.

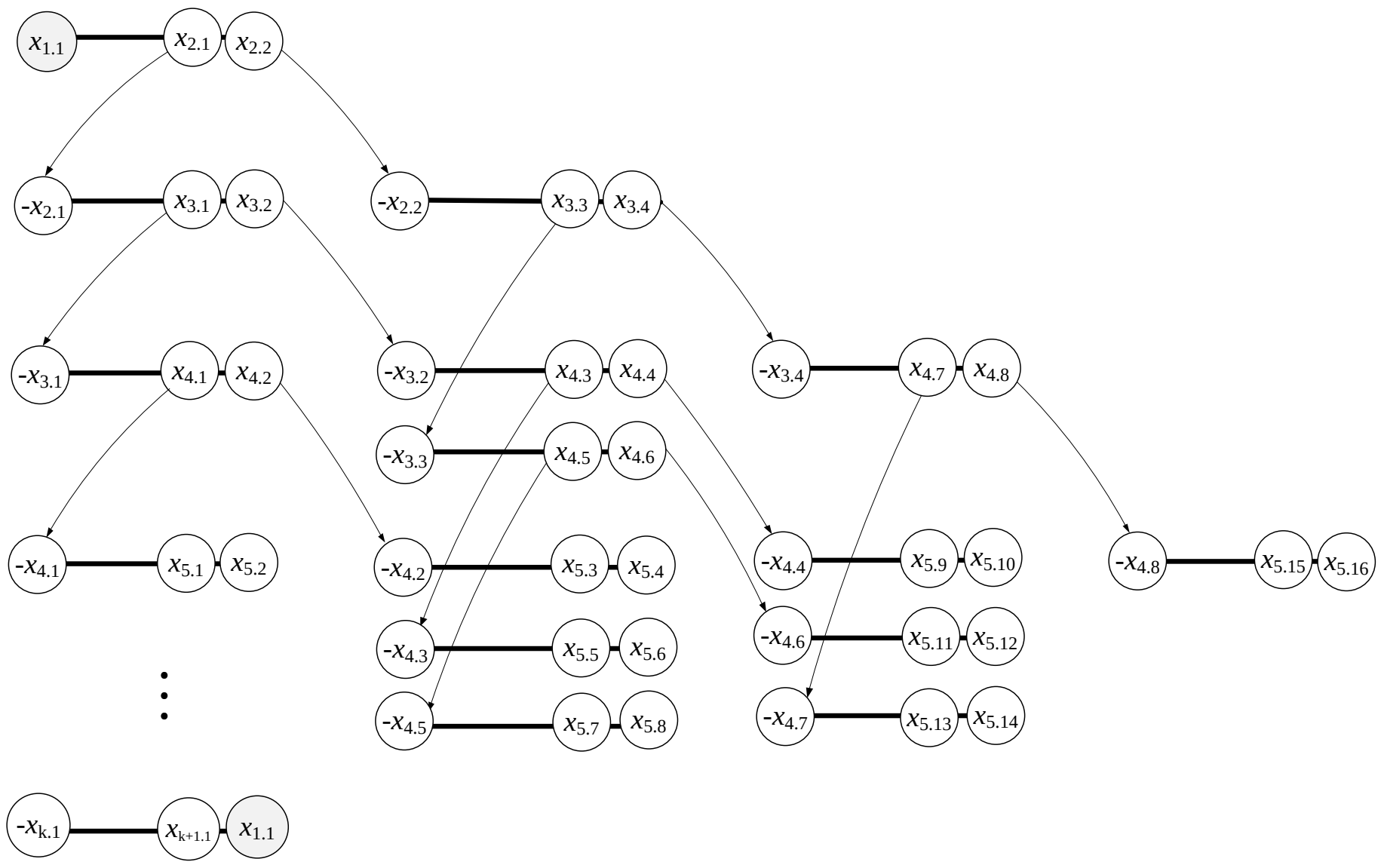

**Fig. 3.** Unit clause transformation into a binary decision tree

Figure 3 shows a decision tree constructed with *k* levels which is generated by the transformation of a unit clause containing variable $x_{1.1}$. The variable $x_{1.1}$ is termed as a **root variable** of the decision tree. All interior nodes have two child nodes and one parent node. Thus this decision tree is termed as a **binary decision tree** constructed with indistinguishable variable pairs. We investigate how deeply the binary decision tree is constructed with *n* input variables. The number of variables needed to construct the tree to the *k-th* level is calculated as

$$\sum_{i=0}^{k} 2^i = n, \quad k = \lfloor \log_2^{n+1} \rfloor -1 \quad \cdots(6)$$

Whenever the level size increases by one, the number of variables needed to construct all nodes of a new level is doubled, which makes the maximum level become a log value of the input size. As the number of selections of a variable to reach a variable in a clause of the *k-th* level becomes *k*, the number of paths to reach a variable in a clause of the *k-th* level becomes $2^k$. We must search for all $2^k$ paths in order to investigate all variables in a clause in the *k-th* level:

$$2^k = 2^{\log_2^{n+1}-1} = \frac{1}{2}\left(n+1\right) \quad \cdots(7)$$

This value is polynomial in input size because the maximum level is a log value of the input size.

## Binomial decision tree of 3-SAT

Now, we investigate how to increase the maximum level of the tree with the same number of input variables in remaining the indistinguishability. Four variables belonging to two indistinguishable variable pairs are different from one another in the case of the binary decision tree. However, there is no way to distinguish two variables in a clause, even if one variable in a clause is the same as the variable contained in another adjacent clause. Therefore, we can replace the adjacent two variables with one variable and reform the binary decision tree as shown in Figure 4.

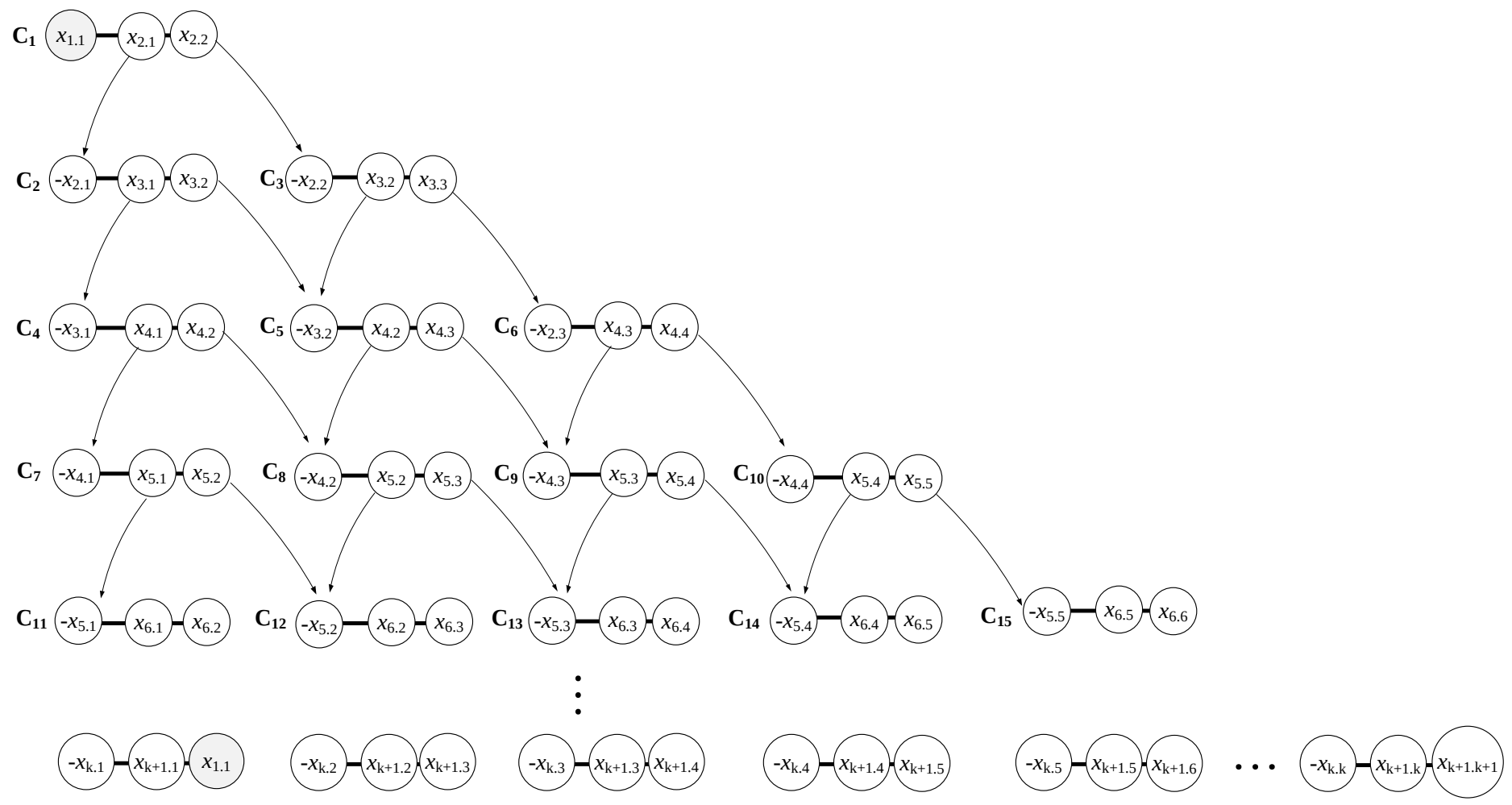

**Fig. 4.** Unit clause transformation into a binomial decision tree

The number of paths starting from the root node and arriving at a child node forms a binomial coefficient. For example, the number of paths arriving at clauses $c_7$, $c_8$, $c_9$, and $c_{10}$ in Figure 4 are ${}_3C_0$, ${}_3C_1$, ${}_3C_2$, and ${}_3C_3$, respectively. Hence, this decision tree is termed as a **binomial decision tree** constructed with indistinguishable variable pairs. In this case, the number of variables needed to construct the tree to the *k-th* level is reduced as

$$\sum_{i=1}^{k+1} i = \frac{(k+1)(k+2)}{2} \quad \cdots(8)$$

The number of selections of a variable needed to reach a variable contained in a clause of the *k-th* level is the same with the level size *k*, which is calculated as

$$\frac{(k+1)(k+2)}{2} = n,\ k = \left\lfloor \frac{\sqrt{8n+1}-3}{2} \right\rfloor \quad (k\text{: level size, } n\text{: input size}) \cdots(9)$$

In the case of a binary decision tree, the maximum level becomes a log value in the input size. However, we can verify that the maximum level of a binomial decision tree becomes a polynomial value in the input size in eq. (9). The binomial decision tree was constructed by the following process. We first transformed a unit clause containing root variable $x_{1.1}$ into a decision chain in order to generate a hard SAT instance and then each variable contained in a clause was divided by two indistinguishable variables in order to increase the hardness. Finally, a one-dimensional decision chain was extended to a two-dimensional decision tree in order to connect two indistinguishable variables to two nodes with the same depth size. As a result, a unit clause containing $x_{1.1}$ is converted to a decision tree containing $x_{1.1}$ twice. We started generating the decision tree in terms of unit clause transformation. However, if we replace the root variable contained in a first node of the last level with any other variable, the dominance property of the root variable disappears because we cannot construct a path to generate clause ($x_{1.1} \vee x_{1.1}$). In addition, the dominance property of the root variable remains, even if we move the root variable contained in a first node of the last level to any other node because we can always create one or more decision paths to generate clause ($x_{1.1} \vee x_{1.1}$). This characteristic of the binomial decision tree is summarized in the following lemma.

**Lemma 1. All variables contained in a binomial decision tree must be investigated in order to decide the satisfiability of an instance.**

*Proof.* If the root variable is contained in another node or two variables in different nodes are replaced with some variable and its complement, then the root variable has dominance property. For example, in Figure 4, if a variable among $x_{6.i}$ ($1 \leq i \leq 6$) is replaced by $x_{1.1}$, then decision paths to generate a unit clause containing variable $x_{1.1}$ are constructed. For example, if $x_{6.2}$ is replaced with $x_{1.1}$, clause sets {$c_1$, $c_2$, $c_4$, $c_7$, $c_{12}$}, {$c_1$, $c_2$, $c_4$, $c_8$, $c_{12}$}, {$c_1$, $c_2$, $c_5$, $c_8$, $c_{12}$}, and {$c_1$, $c_3$, $c_5$, $c_8$, $c_{12}$} generate a new clause ($x_{1.1} \vee x_{1.1}$), by which the dominance property of $x_{1.1}$ is verified. In addition, if $x_{6.3}$ and $x_{6.5}$ are replaced with $z$ and $\neg z$, respectively, six decision paths from $c_1$ to $c_{13}$ generating the clause ($x_{1.1} \vee z$) are constructed and four decision paths from $c_1$ to $c_{14}$ generating the clause ($x_{1.1} \vee \neg z$) are constructed. One more resolution step with ($x_{1.1} \vee z$) and ($x_{1.1} \vee \neg z$) generates a new clause ($x_{1.1} \vee x_{1.1}$), by which the dominance property of $x_{1.1}$ is verified. Therefore, we must investigate all variables contained in a decision tree to verify the dominance property of the root variable. In addition, if a variable has dominance property and the negation of the variable also has dominance property, then the instance is unsatisfiable. Therefore, we must investigate all variables contained in a decision tree to decide the satisfiability of an instance. ■

Figure 5 shows two instances constructed with two binomial decision trees.

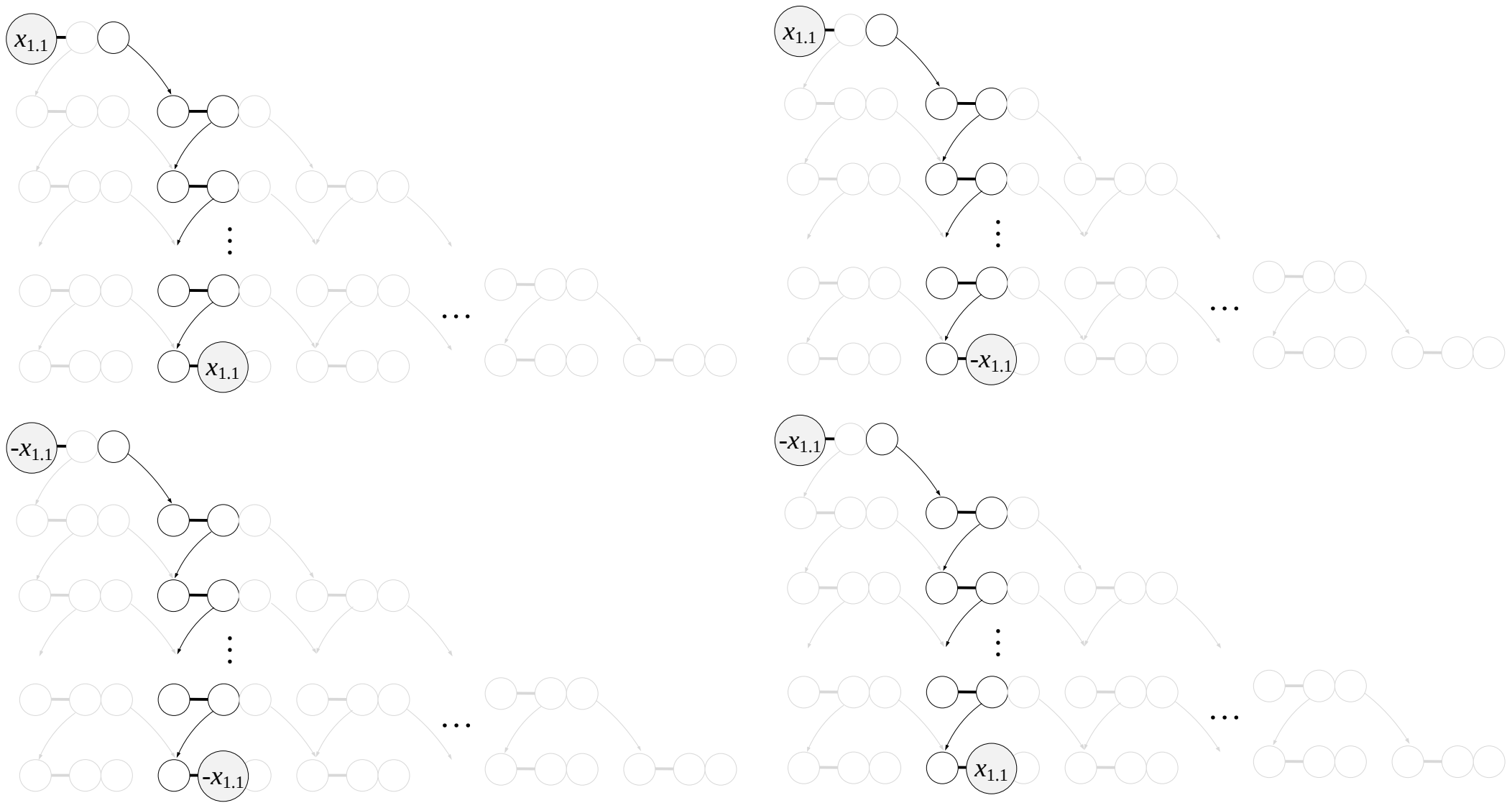

**(a)** **(b)**

**Fig. 5.** Two instances constructed with two binomial decision trees: (a) is unsatisfiable and (b) is satisfiable

There exist one or more decision paths to generate clause ($x_{1.1} \vee x_{1.1}$) in the upper tree and clause ($-x_{1.1} \vee -x_{1.1}$) in the lower tree in Figure 5(a). One resolution step using the clauses ($x_{1.1} \vee x_{1.1}$) and ($-x_{1.1} \vee -x_{1.1}$) generates an empty clause. Therefore, this instance is unsatisfiable. As the locations of variable $x_{1.1}$ and $-x_{1.1}$ are exchanged, two root variables have no dominance property in Figure 5(b). Therefore, the instance is satisfiable.

**Lemma 2. (Row indistinguishability) The decision paths starting from the root node and arriving at child nodes in the same level cannot be distinguished in the binomial decision tree.**

*Proof.* Every decision path is generated by sequential selections of one variable from the indistinguishable variable pair. Therefore, if two paths are equal in size, there is no standard to distinguish each other. ■

**Lemma 3. The number of decision paths starting from the root node and arriving at the clauses of the last level is not polynomially bounded in the binomial decision tree.**

*Proof.* The decision path starting from the root node and arriving at a clause is one-to-one mapped to a clause in the binary decision tree. However, the binomial decision tree has multiple paths arriving at a clause. The number of paths to reach a clause located in the ($k$+1)-*th* level and *i-th* row is calculated as a ($i$-1)-combination of a $k$-element set. The total number of paths to reach a clause of the ($k$+1)-*th* level becomes:

$$\sum_{i=1}^{k+1} \binom{k}{i-1} = 2^k \quad \cdots(10)$$

This sum is equal to the number of paths to reach a variable contained in a clause of the *k-th* level. In addition, we must select one variable between the two at every level. Thus, the number of cases of selections to the *k-th* level also becomes $2^k$. By eq. (9),

$$2^k = 2^{\left\lfloor \frac{\sqrt{8n+1}-3}{2} \right\rfloor} \quad \cdots(11)$$

This value is not polynomially bounded. ■

**Theorem 1. All entry clauses contained in an indistinguishable binomial decision tree cannot be extracted in polynomial time following the decision paths.**

*Proof.* The decision paths are indistinguishable from one another by Lemma 2 (Row indistinguishability). In addition, the number of decision paths is not polynomially bounded by Lemma 3. Therefore, we cannot extract all entry clauses in polynomial time following the decision paths. ■

Now, we investigate whether we can extract all entry causes starting from the root node following the tree levels step by step. This extraction can be executed by constructing tree levels step by step. The clause that can be assigned to a node located in the next level is termed as an **entry candidate clause,** which must contain the negation of the variable used to connect to a child node. First we investigate whether some combinations of entry candidate clauses can be excluded in consideration to construct a tree level.

### Multiple-branching binomial decision tree and redundancy clause

We generated the binomial decision tree from the binary decision tree by imposing the constraint that one variable in a clause is the same with the variable contained in another adjacent clause. If a clause does not contain the same variable with any other clause in a same level, we can construct a new binomial decision tree by assigning the clause as a root node or two binomial decision trees by assigning the negations of the variables contained in the clause as root variables. If all variables are different one another at the *k-th* level, 2*k* numbers of trees connected to the variables can be constructed at maximum. Although the number of trees increases 2*k* times, the number of clauses needed to construct the trees is polynomially bounded because *k* is a polynomial value in the input size.

Figure 6 shows an instance of a multiple-branching binomial decision tree in which all clauses in the third level such as $c_4$, $c_5$, and $c_6$ have different variables from one another. The clauses $c_4$ and $c_6$ play a role of a root node and $c_5$ is connected to two binomial decision trees.

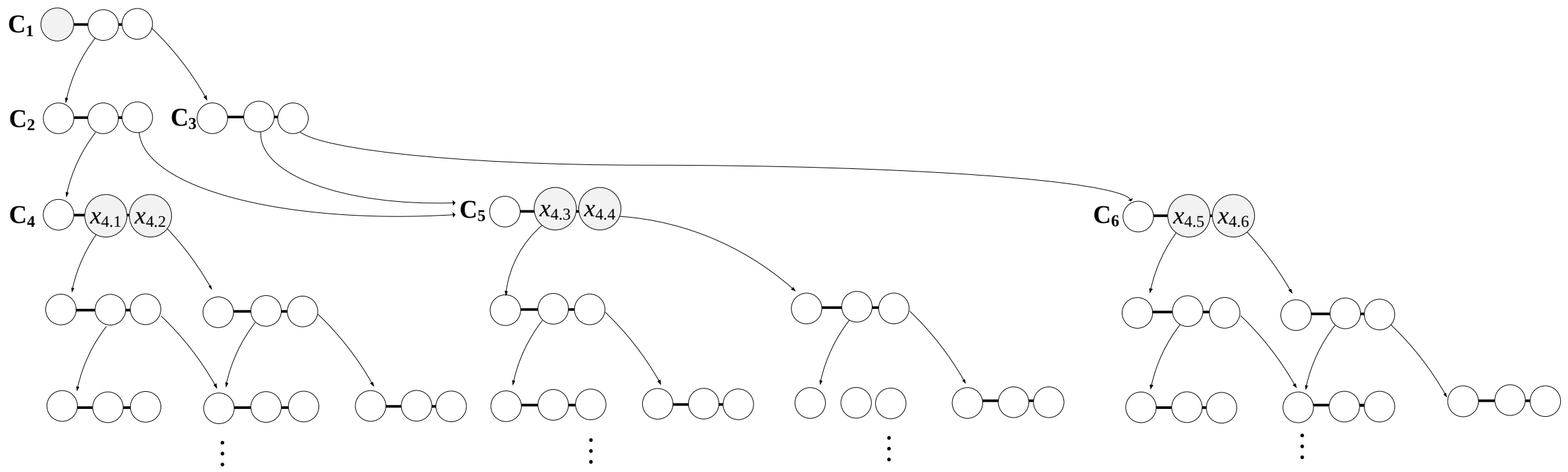

**Fig. 6.** An example of a multiple-branching binomial decision tree

Note that Lemma 1 is also valid in this case. However, any combination of *k* numbers of clauses can be assigned as the nodes at the *k-th* level regardless of whether two adjacent clauses have the same variable. Hence, if an algorithm searches for the clauses constructing the binomial decision tree, two clauses should not be excluded by the reason that the two clauses do not contain the same variable.

Figure 7 shows an example of an implicit decision clause.

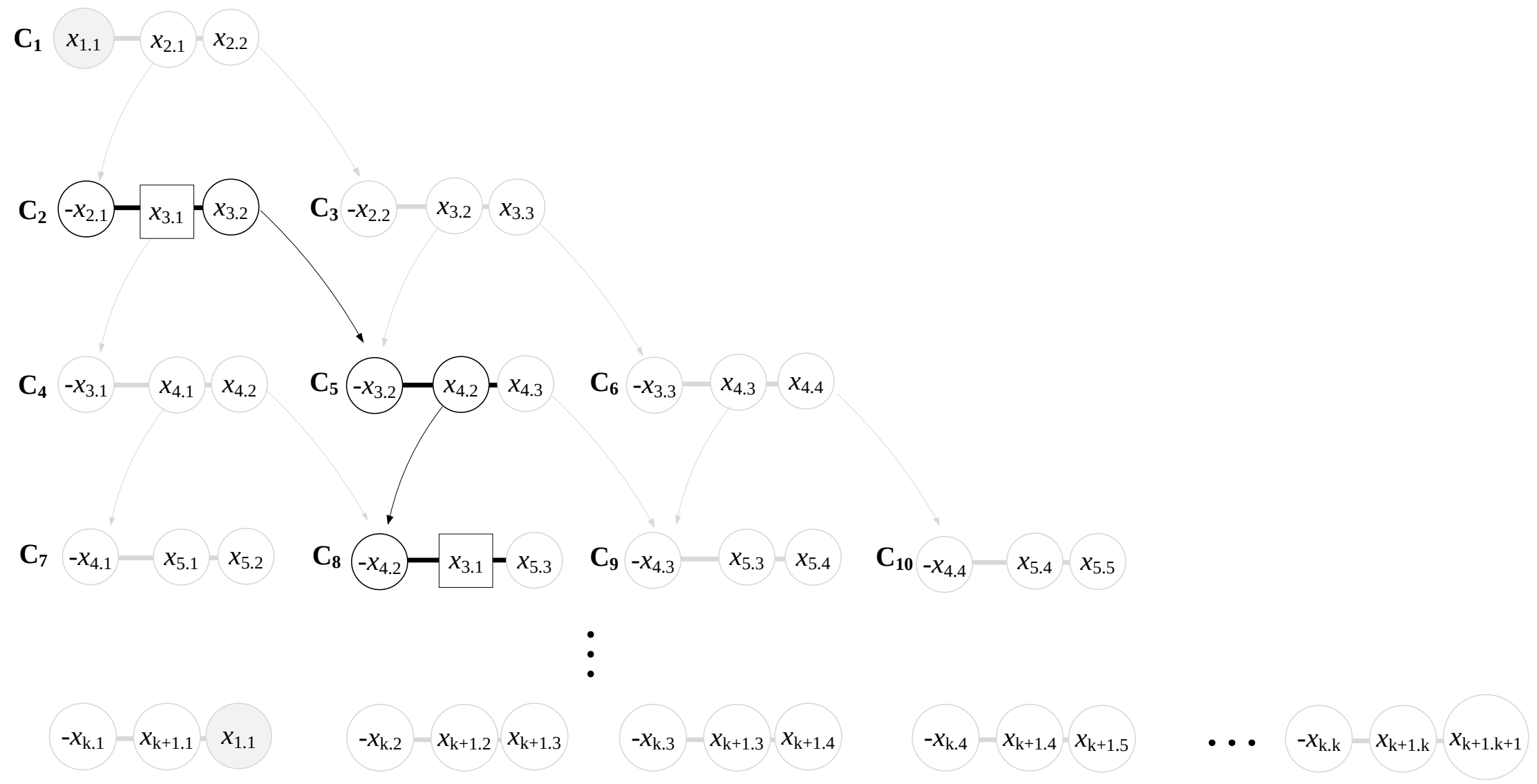

**Fig. 7.** Example of an implicit decision clause

Suppose that all clauses in Figure 7 are explicit decision clauses except for $c_2$. We changed $x_{5.2}$ in $c_8$ with $x_{3.1}$ instead of inserting the switching clause such as $(\neg x_{2.1} \vee x_{3.1} \vee \neg x_{3.2})$ or $(x_{3.1} \vee \neg x_{3.2})$ in the clause set of an instance. We can generate $(\neg x_{3.2} \vee x_{4.2})$ from $c_5$ and $(\neg x_{4.2} \vee x_{3.1})$ from $c_8$ because the clauses $c_5$ and $c_8$ are explicit decision clauses. Thus $(\neg x_{2.1} \vee x_{3.1})$ is generated by clauses $c_2$, $c_5$ and $c_8$. Therefore, $c_2$ becomes an implicit decision clause. If $x_{5.2}$ was not changed to $x_{3.1}$, $c_2$ is not a decision clause because we cannot generate $(\neg x_{2.1} \vee x_{3.1})$. In this case, the dominance property of $x_{1.1}$ cannot be verified because the decision path to generate clause $(x_{1.1} \vee x_{1.1})$ is broken. That is, only $(\neg x_{2.1} \vee x_{3.1} \vee x_{3.2})$ can be used instead of $(\neg x_{2.1} \vee x_{3.1})$. Hence, clause $(x_{1.1} \vee x_{1.1} \vee x_{3.2})$ is generated instead of $(x_{1.1} \vee x_{1.1})$. Note that a SAT instance can contain polynomially many redundancy clauses, which do not affect the satisfiability of the instance regardless of their inclusions. If $(\neg x_{2.1} \vee x_{3.2})$ is generated and $(\neg x_{2.1} \vee x_{3.1})$ is not generated, then, $(\neg x_{2.1} \vee x_{3.1} \vee x_{3.2})$ is always redundancy clause because $x_{3.1}$ can be assigned with any value of '0' or '1'. As a result, if $x_{3.1}$ is contained in any child node, $c_2$ becomes an implicit decision clause. Otherwise, $c_2$ becomes a redundancy clause that is not a decision clause. This means that we must investigate all child nodes in order to verify whether some node clause is an implicit decision clause.

Figure 8 shows an example in which we meet redundancy clauses at the second level and shows another decision tree generated by the choice of the redundancy clauses.

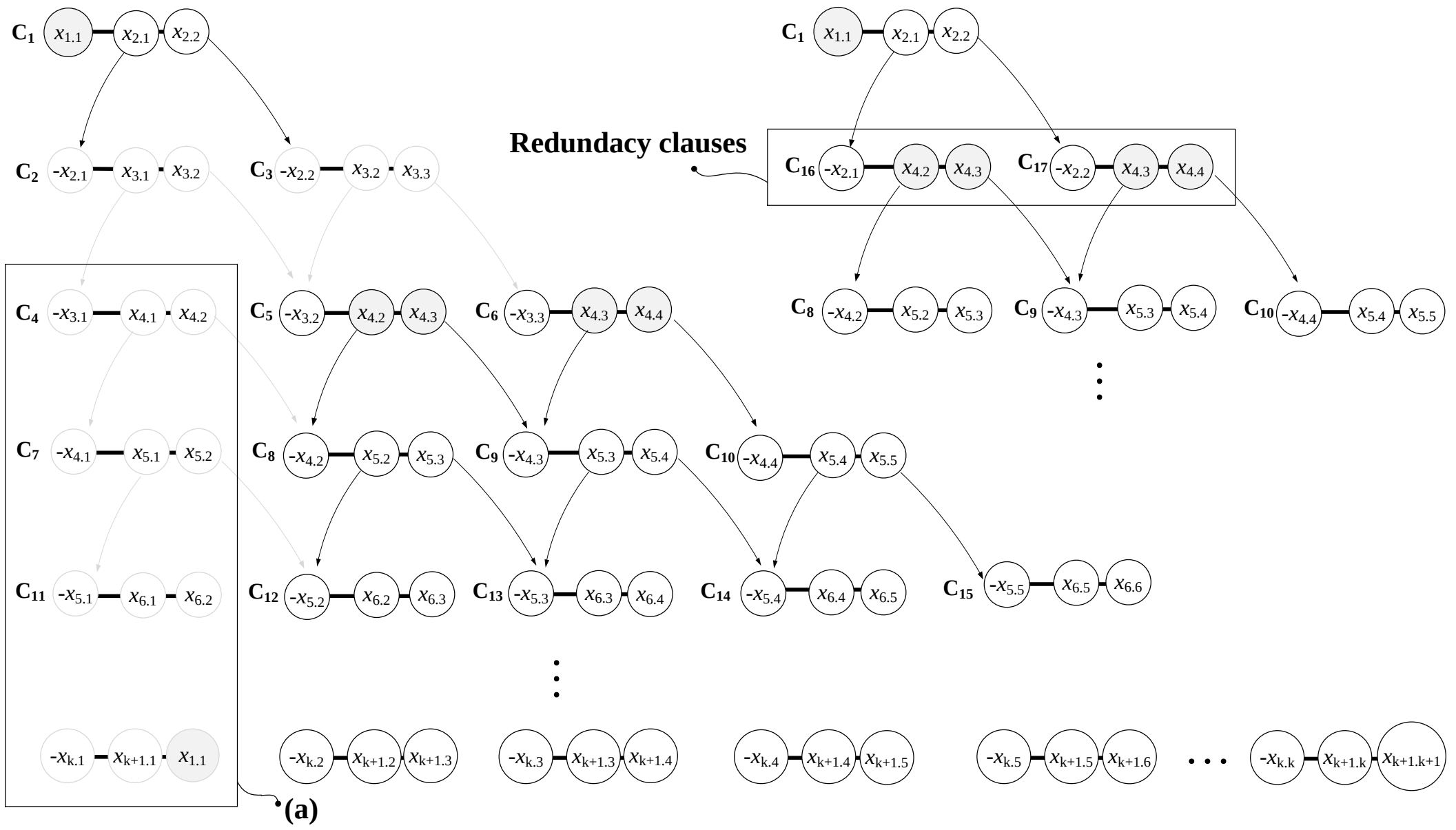

**Fig. 8.** Redundancy clauses that can be assigned as decision clauses

First, we prove that clauses $c_{16}$ and $c_{17}$ are redundancy clauses. We previously mentioned that one node corresponds to three clauses. The clause $(\neg x_{2.1} \lor x_{3.2})$ is generated with $(\neg x_{2.1} \lor x_{3.1} \lor x_{3.2})$ and $(\neg x_{2.1} \lor \neg x_{3.1} \lor x_{3.2})$ by one resolution step and one more resolution step with $(\neg x_{3.2} \lor x_{4.2} \lor x_{4.3})$ generates $c_{16}$:

$$\frac{\neg x_{2.1} \lor x_{3.1} \lor x_{3.2},\ \ \neg x_{2.1} \lor \neg x_{3.1} \lor x_{3.2}}{\neg x_{2.1} \lor x_{3.2}},\quad \frac{\neg x_{2.1} \lor x_{3.2},\ \neg x_{3.2} \lor x_{4.2} \lor x_{4.3}}{\neg x_{2.1} \lor x_{4.2} \lor x_{4.3}}$$

We can generate $c_{17}$ in the same way. As we can generate $c_{16}$ and $c_{17}$ using the clauses constructing the decision tree, two clauses become redundancy clauses. We can easily verify that any variable contained in a child node, can be used when we create a redundancy clause. That is, instead of two indistinguishable variable pair contained in $c_2$, if one or more variables are replaced with any variables contained in child nodes of $c_2$, then a redundancy clause is generated. We can easily create redundancy clauses using this process. Suppose that we selected $c_{16}$ and $c_{17}$ instead of $c_2$ and $c_3$ to construct the second level of the decision tree. Then, the following levels of the decision tree are constructed by the child nodes of $c_5$ and $c_6$. The nodes inside the box (a) as well as $c_2$ and $c_3$ are omitted from the generated decision tree, which make it impossible to verify the dominance property of the root variable. Note that if we assign a redundancy clause to a node, another decision tree is generated which is smaller than the original decision tree. Therefore, in order to extract all entry clauses contained in the decision tree, we must distinguish the entry clause from the redundancy clause. However, there is no standard to distinguish between the entry clause and the redundancy clause. If a redundancy clause is a decision clause, the clause is termed as a **redundancy decision clause**. The decision tree consists of decision clauses, which are termed as **entry decision clauses** in order to distinguish from the redundancy decision clause.

**Lemma 4. (Column indistinguishability) The entry decision clause cannot be distinguished from the redundancy decision clause or the redundancy clause that is not a decision clause.**

*Proof.* The only difference between the entry decision clause and the redundancy decision clause is the locations of the variables contained in the clauses. The variables contained in the entry decision clause are located in a parent node and the variables contained in the redundancy decision clause are located in a child node. We cannot decide which clause is an entry decision clause and which clause is a redundancy decision clause before all entry clauses contained in a decision tree are extracted, which cannot be completed in polynomial time by Theorem 1.

The implicit decision clause must satisfy the condition that one or more child nodes contain the same variable belonging to the implicit decision clause. Therefore, we cannot decide whether a clause is an implicit decision clause before all child nodes of the clause are investigated, which cannot be completed in polynomial time by Theorem 1. ■

**Lemma 5. The number of combinations of entry candidate clauses to construct a tree level of a binomial decision tree is not polynomially bounded.**

*Proof.* As we mentioned above, a SAT instance can contain polynomially many redundancy clauses. Suppose that we constructed an instance in which every node can be assigned with $m$ ($m \geq 2$) clauses including redundancy clauses. At the second level, in order to assign two clauses to two nodes, we need to extract one clause among $m$ clauses and another clause among the other $m$ clauses, which generates $m^2$ combinations of entry candidate clauses. At the *k-th* level, the number of combinations of entry candidate clauses becomes $m^k$. By eq. (9),

$$m^k = m^{\left\lfloor \frac{\sqrt{8n+1}-3}{2} \right\rfloor + 1}, \quad m \geq 2 \quad \cdots (12)$$

This value is not polynomially bounded. ■

**Theorem 2. All entry clauses contained in an indistinguishable binomial decision tree cannot be extracted in polynomial time following the tree levels.**

*Proof.* All combinations of entry candidate clauses can be assigned to the clause set of a tree level regardless of whether two adjacent clauses contain the same variable by considering the multiple-branching binomial decision tree. In addition, the entry decision clause cannot be distinguished from the redundancy decision clause or the redundancy clause that is not a decision clause by Lemma 4. Therefore, we cannot extract only entry decision clauses from the entry candidate clauses. As a result, we must consider all combinations of entry candidate clauses to find the entry decision clauses constructing the tree level. However, the number of combinations of entry candidate clauses to construct a tree level is not polynomially bounded by Lemma 5. Therefore, we cannot extract all entry clauses in polynomial time following the tree levels. ■

**Theorem 3. 3-SAT has no polynomial-time algorithm.**

*Proof*. Suppose that we generated an instance containing one or more multiple-branching binomial decision trees. We must investigate all variables contained in a decision tree to decide the satisfiability of the instance by Lemma 1.

We can extract entry clauses of a decision tree starting from the root node following the decision paths or following the tree levels step by step. However, all entry clauses contained in an indistinguishable binomial decision tree cannot be extracted in polynomial time following the decision paths by Theorem 1. In addition, all entry clauses contained in an indistinguishable binomial decision tree cannot be extracted in polynomial time following the tree levels by Theorem 2. Therefore, we cannot decide the satisfiability of the instance in polynomial time because we cannot investigate all entry clauses of the decision tree in polynomial time. As a result, 3-SAT has no polynomial-time algorithm. ■

**Theorem 4. Class P is a proper subset of class NP.**

*Proof*. Any deterministic Turing machine can be simulated by a non-deterministic Turing machine with no overhead. Thus, class P is included in class NP. 3-SAT is included in class NP[2]. However, 3-SAT is not included in class P by Theorem 3. Therefore, P $\subsetneq$ NP ■

# DISCUSSION

Row indistinguishability ($P_1$ and $P_2$ in Figure 9) and column indistinguishability ($C_1$ ~ $C_6$ in Figure 9) are caused by indistinguishable variable pairs and redundancy clauses, respectively.

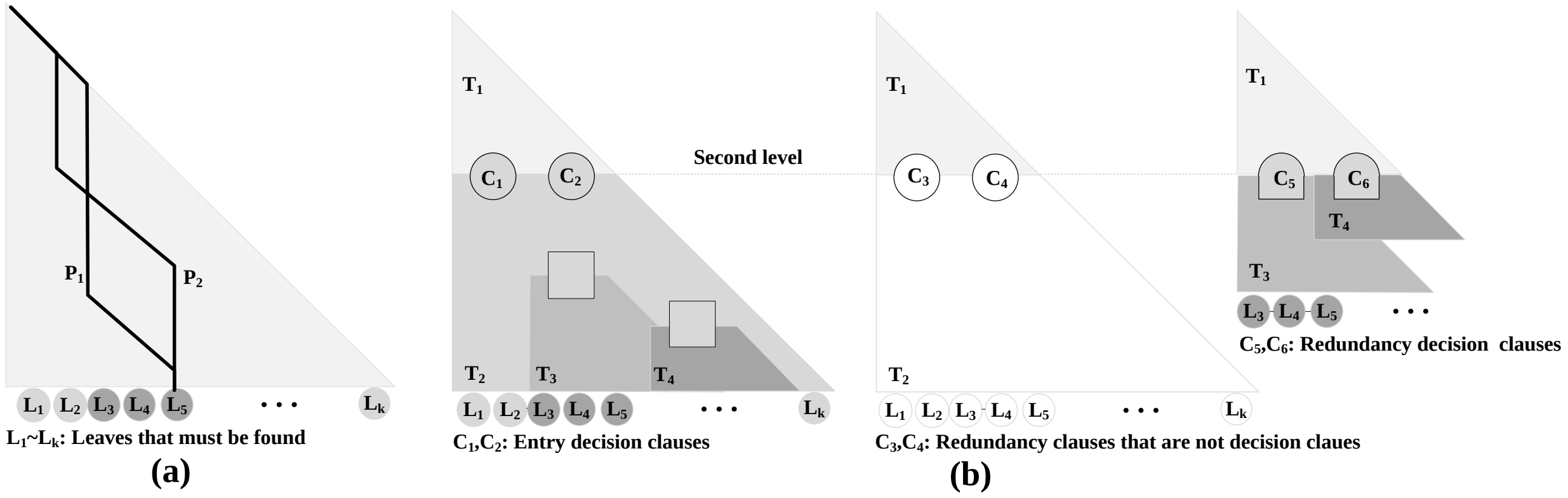

**Fig. 9.** Conceptual diagram of row indistinguishability (a) and column indistinguishability (b) of an indistinguishable binomial decision tree

Row indistinguishability and exponentially many decision paths make it impossible to search for all entry clauses in polynomial time following the vertical decision paths. In addition, column indistinguishability and exponentially many combinations of entry candidate clauses make it impossible to search for all entry clauses in polynomial time following the horizontal tree levels. This result leads to the conclusion that 3-SAT has no polyn3omial-time algorithm.

The non-existence of a polynomial-time algorithm for 3-SAT immediately leads to the conclusion that class P is a proper subset of class NP. The binomial decision tree is constructed with polynomial number of clauses which make it possible to verify the solution in polynomial time. However, the binomial decision tree generates exponentially many indistinguishable paths which make it impossible to solve the problem in polynomial time. Hence, the indistinguishable binomial decision tree clearly explains the relationship of verifiability and solvability. Every NP-complete problem is reduced to intractable SAT in polynomial time. In addition, every intractable SAT is reduced to 3-SAT in polynomial time. Hence, every NP-complete problem's relationship of verifiability and solvability is definitely explained by the characteristic of the indistinguishable binomial decision tree.

## MATERIALS AND METHODS

### Resolution technique to solve the SAT problem

The procedure to decide the satisfiability of an instance by a sequence of resolution steps is as follows. First, we create a clause set *S* with all clauses in the CNF instance. We apply the resolution rule to all possible pairs of clauses that contain complementary literals. After each application of the resolution rule, we simplify the resulting clause by removing repeated literals. If the clause contains complementary literals, it is discarded. Otherwise, if the resulting clause is not yet present in the clause set *S*, the resulting clause is added to *S*, and is considered for further resolution steps. After applying a resolution rule, if the empty clause is derived, we decide that the instance is unsatisfiable. On the other hand, if the empty clause cannot be derived, and the resolution rule cannot be applied to derive any more new clauses, which is said to be saturated, we decide that the instance is satisfiable.